\documentclass[aps,prc,preprint,amsmath,amssymb,showpacs,preprintnumbers,superscriptaddress]{revtex4-1}
\usepackage{CJK}
\usepackage{graphicx}
\usepackage{dcolumn}
\usepackage{bm}
\usepackage{color}
\usepackage{hyperref}
\allowdisplaybreaks[4]

\begin{document}
\begin{CJK*}{UTF8}{}

  \title{Nuclear level density from relativistic density functional theory and combinatorial method}
  \author{X. F. Jiang \CJKfamily{gbsn} (姜晓飞)}
  \affiliation{State Key Laboratory of Nuclear Physics and Technology, School of Physics, Peking University, Beijing 100871, China}

  \author{X. H. Wu \CJKfamily{gbsn} (吴鑫辉)}
  \affiliation{Department of Physics, Fuzhou University, Fuzhou 350108, Fujian, China}

  \author{P. W. Zhao \CJKfamily{gbsn} (赵鹏巍)}
  \affiliation{State Key Laboratory of Nuclear Physics and Technology, School of Physics, Peking University, Beijing 100871, China}

  \author{J. Meng \CJKfamily{gbsn} (孟杰)}
  \email{mengj@pku.edu.cn}
  \affiliation{State Key Laboratory of Nuclear Physics and Technology, School of Physics, Peking University, Beijing 100871, China}
  \affiliation{Yukawa Institute for Theoretical Physics, Kyoto University, Kyoto 606-8502, Japan}

  \begin{abstract}
    Nuclear level density is calculated with the combinatorial method based on the relativistic density functional theory including pairing correlations.
    The Strutinsky method is adopted to smooth the total state density in order to refine the prediction at low excitation energy.
    The impacts of pairing correlations and moments of inertia on the nuclear level density are discussed in detail.
    Taking $\mathrm{^{112}Cd}$ as an example, it is demonstrated that the nuclear level density based on the relativistic density functional PC-PK1 can reproduce the experimental data at the same level as or even better than the previous approaches.
 \end{abstract}


  \maketitle

\end{CJK*}

\section{Introduction}\label{Section1}

The origin of elements is one of the most fundamental scientific problems.
Currently, the rapid neutron capture process (\emph{r}-process) is believed to be responsible for the nucleosynthesis of about half of the elements heavier than Iron~\cite{Burbidge1957RMP, Kajino2019PPNP, Cowan2021RMP}.
In \emph{r}-process simulations, neutron capture rates of thousands of neutron-rich nuclei are required, and most of these nuclei lie far beyond the experimentally known region.
Therefore, the \emph{r}-process models rely on the theoretical predictions for the required neutron capture rates~\cite{Larsen2019PPNP}.
At present, neutron-capture rates are usually calculated with the Hauser-Feshbach model~\cite{Hauser1952PR}, in which the nuclear level density plays a crucial role.

Accurate prediction of the nuclear level density is challenging because of the exponential growth of the nuclear level density with the excitation energy, and the complex nuclear structure and dynamics such as shell structures, pairing correlations, and collective motions~\cite{Ericson1960AP, NuclearStructure1}.
In a pioneering work~\cite{Bethe1936PR}, Bethe proposed a level density formula based on the Fermi gas model with equidistant spacing of the single-particle levels near the Fermi level, in which pairing correlations and collective excitations were not considered.
To include the pairing correlations, the conventional shifted Fermi gas model~\cite{Newton1956CJP, Cameron1958CJP} is proposed by shifting the excitation energy according to the semiempirical pairing energies.
By treating the energy shift as a free parameter, the back-shifted Fermi gas model~\cite{Vonach1968NPA, Huizenga1969PR, Vonach1969NPA} leads to a better description of the level density at low excitation energy.
By using an empirical constant temperature formula, the Gilbert-Cameron model~\cite{Gilbert1965CJPb} also improves the nuclear level density at low excitation energy.
For better consideration of the pairing correlations, the Bardeen-Cooper-Schrieffer (BCS) theory is adopted in the generalized superfluid model~\cite{Ignatyuk1979SJNP, Ignatyuk1993PRC}, in which the collective excitations are taken into account with the enhancement factors.
By solving a pairing Hamiltonian, along with the single-particle levels in a Woods-Saxon well plus a spin-orbit interaction, the partition function and nuclear level density are calculated in Ref.~\cite{Hung2017PRL, Dang2017PRC}.

The conventional nuclear shell model exactly solves a many-body Hamiltonian within a given model space, and could in principle provide an exact prediction of the nuclear level density.
However, the computational cost is highly demanding.
The shell-model Monte Carlo approach~\cite{Nakada1997PRL, Alhassid1999PRL, Alhassid2007PRL, Alhassid2015PRC} solves this problem and has been applied to nuclei as heavy as lanthanides~\cite{Alhassid2008PRL, Ozen2013PRL, Ozen2015PRC}, but it is limited to schematic nuclear Hamiltonian due to the sign problem.
More realistic interactions can be employed in the moment method~\cite{Sen'kov2010PRC, Sen'kov2013CPC, Sen'kov2016PRC}, the stochastic estimation method ~\cite{Shimizu2016PLB, Chen2023PRC}, and the extrapolated Lanczos matrix approach~\cite{Ormand2020PRC} but are limited to light and medium-mass nuclei only~\cite{Fanto2021PRC}.
Self-consistent mean-field methods can be used to describe nuclear level density across the entire nuclear chart.
Based on nuclear properties provided by mean-field calculations, the nuclear level density is usually calculated using either the statistical method or the combinatorial method.
Many calculations of this kind have been reported, including the extended Thomas-Fermi approximation with Skyrme forces plus statistical method~\cite{Kolomietz2018PRC}, the Skyrme Hartree-Fock-BCS plus statistical method~\cite{Demetriou2001NPA, Minato2011JNST}, the Skyrme Hartree-Fock-Bogoliubov plus combinatorial method~\cite{Goriely2008PRC}, and the temperature-dependent Gogny Hartree-Fock-Bogoliubov plus combinatorial method~\cite{Hilaire2012PRC}.
More recently, to better include correlations beyond the mean field, a new method based on the boson expansion of QRPA excitations has been proposed~\cite{Hilaire2023PLB}.

During the past decades, the relativistic density functional theory (DFT) has been successfully applied to describe both ground-state and excited properties of atomic nuclei~\cite{Mengbook}.
For nuclear level density, the relativistic DFT has been used to calculate the level density with both the combinatorial method~\cite{GengKP2023NST} and the statistical method~\cite{Zhao2020PRC}.
It should be emphasized that the differences between the statistical method and the combinatorial method are fundamental.
In the statistical method, also known as the Darwin-Fowler method and partition function method, the intrinsic level density is obtained with an inverse Laplace transform of the partition function with the saddle-point approximation~\cite{Ericson1960AP, NuclearStructure1}, and the collective excitations are taken into account by enhancement factors.
In the combinatorial method, however, the states of single-particle excitations are exactly counted by expanding a generating function constructed from the single-particle levels.
The combinatorial method provides the energy-, spin-, and parity-dependent nuclear level densities and describes their nonstatistical behaviors~\cite{Goriely2022PRC} which significantly influence the cross section predictions~\cite{Goko2006PRL}.
Non-relativistic calculations with the combinatorial method have been proposed~\cite{Hilaire2001EPJA, Hilaire2006NPA, Goriely2008PRC, Hilaire2012PRC} and the predictions have been used as input data in the TALYS code package~\cite{Koning2012NDS}.
The relativistic DFT calculations with the combinatorial method have been reported in Ref.~\cite{GengKP2023NST}, while the pairing correlations are neglected.
Furthermore, to remove the large fluctuations of state density at low excitation energy, a smoothing procedure is required.
However, the conventional smoothing method described in Ref.~\cite{Hilaire2001EPJA} faces difficulties in correctly describing total state densities at low excitation energy, leading to unsatisfactory reproductions of the corresponding experimental data.
Therefore, it is necessary to improve the combinatorial method with a proper smoothing method.

In this work, we calculate the nuclear level density with the combinatorial method based on the relativistic DFT.
The relativistic Hartree-Bogoliubov (RHB) theory is employed to calculate nuclear properties.
The pairing correlations ignored in the previous investigation are taken into account at the mean-field level and in single-particle excitations as well.
A more mathematically appropriate method, the Strutinsky method, is adopted to remove the large fluctuations of state density at low excitation energy.
The nucleus $\mathrm{^{112}Cd}$ is taken as an example to show the results.

\section{Theoretical Framework}\label{Section2}

\subsection{The relativistic Hartree-Bogoliubov theory}\label{Subsection2.1}

The relativistic Hartree-Bogoliubov (RHB) theory is employed to provide a unified and self-consistent treatment of mean field and pairing correlations.
The detailed formulism of the RHB theory can be seen in Refs.~\citep[]{Ring1996PPNP, Vretenar2005PR, Meng2006PPNP, Niksic2011PPNP}.
In the RHB theory, one needs to solve self-consistently the RHB equation for nucleons,
\begin{equation}
  \left(\begin{array}{c c}{{\hat{h}_{\mathrm{D}}-\lambda_\tau}}&{{\hat{\Delta}}}\\ {{-\hat{\Delta}^{\ast}}}&{{-\hat{h}_{\mathrm{D}}^{\ast}+\lambda_\tau}}\end{array}\right)\left(\begin{array}{c}{{U_{k}}}\\ {{V_{k}}}\end{array}\right)=E_{k}\left(\begin{array}{c}{{U_{k}}}\\ {{V_{k}}}\end{array}\right),\label{Eq1}
\end{equation}
where $\hat{h}_{\mathrm{D}}$ is the single-nucleon Dirac Hamiltonian, $\lambda_\tau$ is the Fermi energy ($\tau=n/p$ for neutrons and protons), $\hat{\Delta}$ is the pairing potential, $U_k$ and $V_k$ are the quasi-particle wavefunctions, and $E_k$ is the corresponding quasi-particle energy.
The single-nucleon Dirac Hamiltonian $\hat{h}_{\mathrm{D}}$ reads
\begin{equation}
  \hat{h}_{\mathrm{D}}=\bm\alpha\cdot\bm p+\beta(m+S)+V,
\end{equation}
where,
\begin{align}
  S & =\alpha_{S}\rho_{S}+\beta_{S}\rho_{S}^{2}+\gamma_{S}\rho_{S}^{3}+\delta_{S}\Delta\rho_{S},                                                    \\
  V & =\alpha_{V}\rho_{V}+\gamma_{V}(\rho_{V})^{3}+\delta_{V}\Delta\rho_{V}+\tau_{3}\alpha_{TV}\rho_{TV}+\tau_{3}\delta_{TV}\Delta\rho_{TV}+eA^{0}.
\end{align}
The pairing potential $\hat\Delta$ is determined by
\begin{equation}
  \hat\Delta_{ab}=\frac{1}{2}\sum_{cd}\langle ab|V^{pp}|cd\rangle_{a}\kappa_{cd},
\end{equation}
where $V^{pp}$ is the pairing force and $\kappa$ is the pairing tensor.
The RHB equation~\eqref{Eq1} needs to be solved self-consistently because, the scalar, vector, and pairing potentials are coupled self-consistently with the densities and, in turn, the quasi-particle wavefunctions.

\subsection{The combinatorial method}\label{Subsection2.2}

In the combinatorial method, the nuclear level density is calculated based on nuclear single-particle levels, masses, radii, deformations, and moments of inertia, and the detailed formulism can be seen in Refs.~\cite[]{Hilaire2001EPJA, Hilaire2006NPA, Goriely2008PRC, Hilaire2012PRC}.
The single-particle levels are used to calculate the number of incoherent particle-hole (ph) states as functions of the single-particle excitation energy, the angular momentum projection onto the symmetry axis, and the parity.
For a given excitation energy of the nucleus, the total number of the states is deduced by folding the incoherent ph states and the collective vibration states whose number is counted by introducing a generalized boson partition function~\cite[]{Goriely2008PRC}.
In principle, one can derive the total state density at any excitation energy by counting the total number of states.
However, one should keep in mind that the excitation energies are continuous so there are infinitely many of them.
Therefore, the numbers of the folded states are counted only at a series of equally spaced excitation energies which are expressed as integer multiples of an energy discretization unit $\varepsilon_0$.
The resultant total state densities $\rho$ at each excitation energy are then given by
\begin{equation}
  \rho(U,K,P)=\frac{C(U,K,P)}{\varepsilon_0}.\label{eq6}
\end{equation}
Here, $U$ is the excitation energy, $K$ is angular momentum projection onto the symmetry axis, $P$ is parity, and $C(U,K,P)$ is the number of the folded states in an energy interval $\varepsilon_0$ centered around $U$.
It turns out that the obtained total state densities $\rho$ in Eq.~\eqref{eq6} strongly depend on the energy discretization unit $\varepsilon_0$, and one cannot obtain a smooth total state density as a function of the excitation energy $U$ even at very small $\varepsilon_0$.
Consequently, a smoothing method is required to obtain smooth total state densities against the excitation energy.

In the conventional smoothing method~\cite[]{Hilaire2001EPJA}, the logarithm of the cumulated number of states is first obtained with
\begin{equation}
  \ln N(U,K,P)=\ln\left[\sum_{U^{\prime}=0}^{U}C(U^{\prime},K,P)\right],
\end{equation}
and then the total state densities are obtained with
\begin{equation}
  \tilde{\rho}(U,K,P)\equiv\frac{dN(U,K,P)}{dU}=N(U,K,P)\frac{d\ln N(U,K,P)}{dU},\label{eq7}
\end{equation}
where $N(U,K,P)$ and $d\ln N(U,K,P)/dU$ are deduced from a linear interpolation, over an energy interval $\delta U$ centered around $U$, of the $\ln N(U,K,P)$ values.
This method gives a smooth total state density against the excitation energy, but is found to present poor results at low excitation energy.

To improve the predictions at low excitation energy, the Strutinsky method~\citep[]{Strutinsky1967NPA, Strutinsky1968NPA, Strutinsky1975NPA, ManyBody, NiuYF2009CPL}, is adopted to directly smooth the total state density $\rho$ given by Eq.~\eqref{eq6} with
\begin{equation}
  \tilde{\rho}(U,K,P)=\frac{1}{\gamma_0}\int_{-\infty}^{+\infty}\rho(U^{\prime},K,P)f\left (\frac{U^{\prime}-U}{\gamma_0}\right )dU^{\prime},\label{Strutinsky}
\end{equation}
where the function $f(x)$ is constructed as $f(x)=\frac{1}{\sqrt{\pi}}e^{-x^2}P(x)$,
where $P(x)$ is a generalized Laguerre polynomial $L_{M_0}^{1/2}\left(x^2\right)$~\citep[]{Abramowitz1965Handbook}.
Here, $\gamma_0$ is the smoothing range, and $M_0$ is the order of the generalized Laguerre polynomial.
The Strutinsky method can be justified with two important facts.
Firstly, it would keep the smoothed function unchanged if smoothed again with the same procedure.
Secondly, it strictly fulfills the conservation of the number of states.
The second fact is essential for the accuracy of the calculation, especially at low excitation energy.

Based on the obtained smooth total state densities, the nuclear level densities can be deduced.
For a deformed nucleus, collective rotation effects must be included, by building up rotational bands consistently on each of the folded states.
The nuclear level density is, thus, given by~\cite[]{Hilaire2001EPJA}:
\begin{equation}
  \begin{split}
    \rho_\text{def}(U,J,P)=&\frac{1}{2}\left[\sum_{K=-J,K\neq0}^J\tilde{\rho}(U-E_\text{rot}^{J,K},K,P)\right]\\
    &+\tilde{\rho}(U-E_\text{rot}^{J,0},0,P)
    \left[\delta_{(J=\text{even})}\delta_{(P=+)}+\delta_{(J=\text{odd})}\delta_{(P=-)}\right],\label{nld}
  \end{split}
\end{equation}
where $\tilde{\rho}$ is the smooth total state density, $U$ is the excitation energy, $J$ is angular momentum, $K$ is angular momentum projection onto the symmetry axis, and $P$ is parity.
The $\delta_{(x)}$ equals to $1$ if $x$ holds true and $0$ otherwise, which restricts the rotational bands established on $K=0$ states to the level sequences $J=0,2,4,...$ for $P=+$, and $J=1,3,5,...$ for $P=-$.
The rotation energy $E_\text{rot}^{J,K}$ is obtained with
\begin{equation}
  E_{\text{rot}}^{J,K}=\frac{\hbar^2}{2\mathcal{J}_{\perp}}\left[J(J+1)-K^2\right],\label{rotenergy}
\end{equation}
where $\mathcal{J}_{\perp}$ is the moment of inertia of a nucleus rotating around an axis perpendicular to the symmetry axis.

Three formulas of moments of inertia (MOI) are examined in the work, i.e., the MOI of a rigid rotor
\begin{equation}
  \mathcal{J}_\zeta^\text{rig}=\frac{2}{5}MAR^2\left[1-\sqrt{\frac{5}{4\pi}}\beta_2\cos\left(\gamma-\frac{2\pi}{3}\zeta\right)\right],\label{rigmoi}
\end{equation}
the MOI of irrotational fluid
\begin{equation}
  \mathcal{J}_\zeta^\text{irr}=\frac{3}{2\pi}MAR^2\beta_2^2\sin^2\left(\gamma-\frac{2\pi}{3}\zeta\right),\label{irrmoi}
\end{equation}
and the microscopic Inglis-Belyaev formula~\citep[]{Inglis1956PR, Beliaev1961NP}
\begin{equation}
  \mathcal{J}_\zeta^\text{ing}=\sum_{i,j}\frac{(u_iv_j-v_iu_j)^2}{E_i+E_j}|\langle i|\hat{J_\zeta}|j\rangle|^2.\label{ingmoi}
\end{equation}
Here, $A$ is the mass number, and $\zeta$ denotes the axis of rotation.
In Eq.~\eqref{rigmoi} and~\eqref{irrmoi}, the nuclear mass $M$, quadrupole deformation parameter $\beta_2$, and nuclear radius $R$ are determined by solving the RHB equation.
In Eq.~\eqref{ingmoi}, the occupation probabilities $v_i$, and single-particle wavefunctions $\psi_i$ are obtained from quasi-particle wavefunctions by canonical transforming~\citep[]{ManyBody}.
The energies in Eq.~\eqref{ingmoi} are $E_i=\sqrt{(\epsilon_i-\lambda)^2+\Delta_i^2}$, in which $\lambda$ is Fermi energy, $\epsilon_i$ is single-particle energy, and $\Delta_i$ is energy gap.
The summation in Eq.~\eqref{ingmoi} runs over the proton and neutron single-particle states.

\section{Numerical details}\label{Section3}

In the present work, the relativistic density functional PC-PK1~\cite{Zhao2010PRC} is employed in the RHB calculation and, in the pairing channel, a finite-range separable pairing force with the pairing strength $G=728\ \mathrm{MeV\ fm^3}$~\cite{Tian2009PLB} is adopted.
The RHB equation is solved by expanding the quasi-particle wavefunctions in terms of a three-dimensional harmonic oscillator basis in Cartesian coordinates~\cite{Niksic2014CPC}, which contains 14 major shells.
The obtained ground-state deformation of $\mathrm{^{112}Cd}$ is $\beta_2=0.145$.

In the calculation of nuclear level density, the cut-off of the excitation energy and angular momentum are taken as $10\ \mathrm{MeV}$ and $49\hbar$ respectively, and the energy discretization unit is taken as $\varepsilon_0=0.005, 0.01, 0.05 \ \mathrm{MeV}$ respectively.
The energy interval in the conventional smoothing method is $\delta U=0.2\ \mathrm{MeV}$~\citep[]{Hilaire2001EPJA}.
The parameters in the Strutinsky method are chosen to be $\gamma_0 = 0.2\ \mathrm{MeV}$ and $M_0=1$.

\section{Results and discussion}\label{Section4}

\begin{figure}[!htbp]
  \centering
  \includegraphics[width=0.4\textwidth]{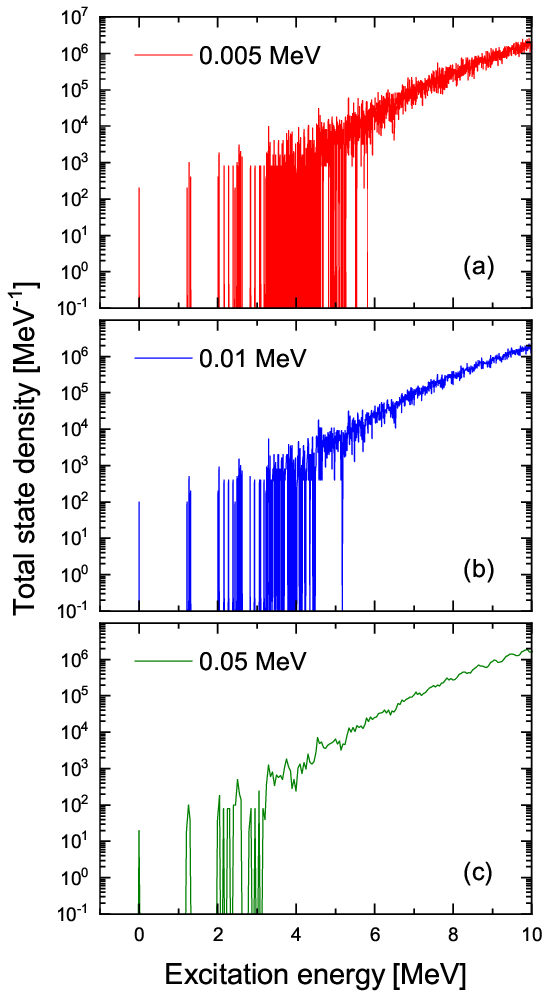}
  \caption{(color online) The total state densities $\rho$ of $\mathrm{^{112}Cd}$ with energy discretization units $\varepsilon_0$ taken as $0.005$ (a), $0.01$ (b), and $0.05\ \mathrm{MeV}$ (c) respectively for comparison.}
  \label{fig1}
\end{figure}

The dependence of the total state densities $\rho$ given by Eq.~\eqref{eq6} on the energy discretization unit $\varepsilon_0$ in the combinatorial method is illustrated in Fig.~\ref{fig1}, where the total state densities $\rho$ of $\mathrm{^{112}Cd}$ are shown as functions of excitation energies.
A dramatic dependency on $\varepsilon_0$ is found, i.e., the smaller $\varepsilon_0$, the more significant the fluctuations.
It could be foreseen that if $\varepsilon_0$ were to be reduced to zero, the $\rho$ would become a combination of a series of delta functions lying on the exact excitation energies of the folded states.
It is therefore understandable that, in Eq.~\eqref{eq6}, a small $\varepsilon_0$ does not lead to smooth total state densities. Consequently, a smoothing method is required to remove the large fluctuations and obtain smooth total state densities.

\begin{figure}[!htbp]
  \centering
  \includegraphics[width=0.4\textwidth]{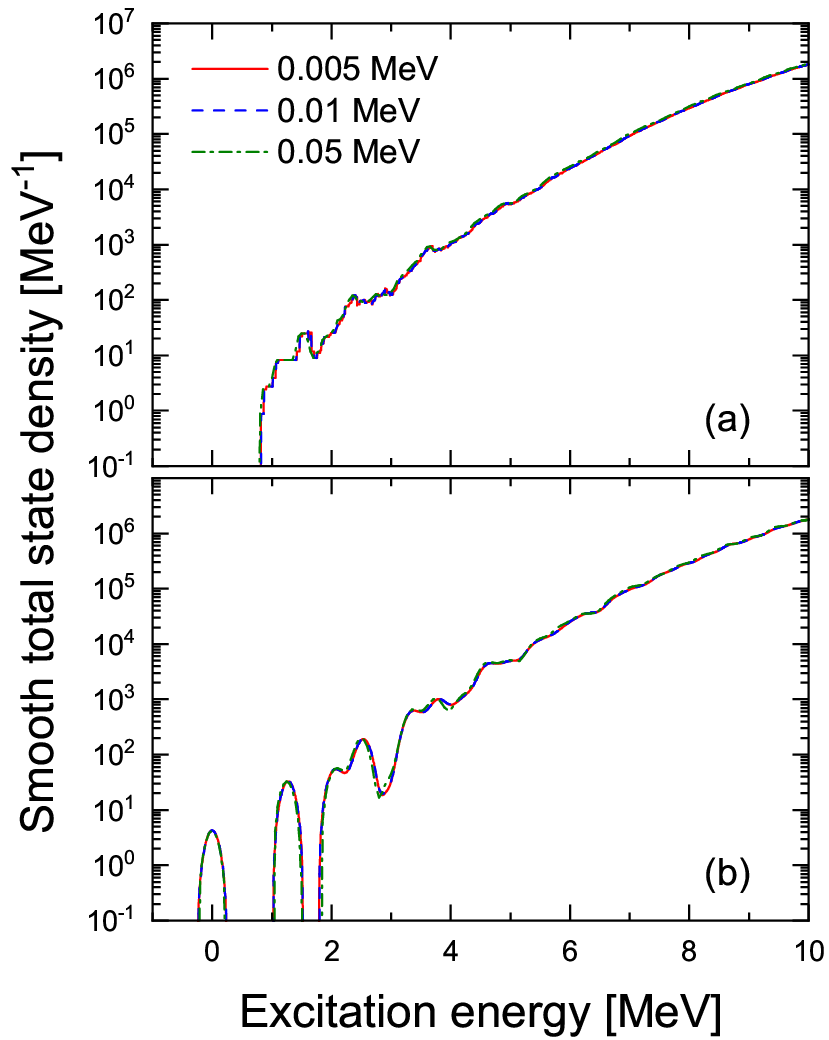}
  \caption{(color online) The smooth total state densities $\tilde{\rho}$ of $\mathrm{^{112}Cd}$ obtained with the conventional smoothing method (a) and the Strutinsky method (b).}
  \label{fig2}
\end{figure}

The smooth total state densities $\tilde{\rho}$, obtained with the conventional smoothing method and the Strutinsky method, are shown in Fig.~\ref{fig2}.
The total state densities $\rho$ are significantly smoothed by both methods.
However, the Strutinsky method better presents the details at low excitation energy, in particularly the ground state.
This would influence the nuclear level density.
Moreover, smooth total state densities $\tilde{\rho}$ barely change when $\varepsilon_0$ is taken from $0.005\ \mathrm{MeV}$ to $0.01\ \mathrm{MeV}$, and even to $0.05\ \mathrm{MeV}$.
Therefore, in the present work, we take $\varepsilon_0=0.01\ \mathrm{MeV}$.

\begin{figure}[!htbp]
  \centering
  \includegraphics[width=0.4\textwidth]{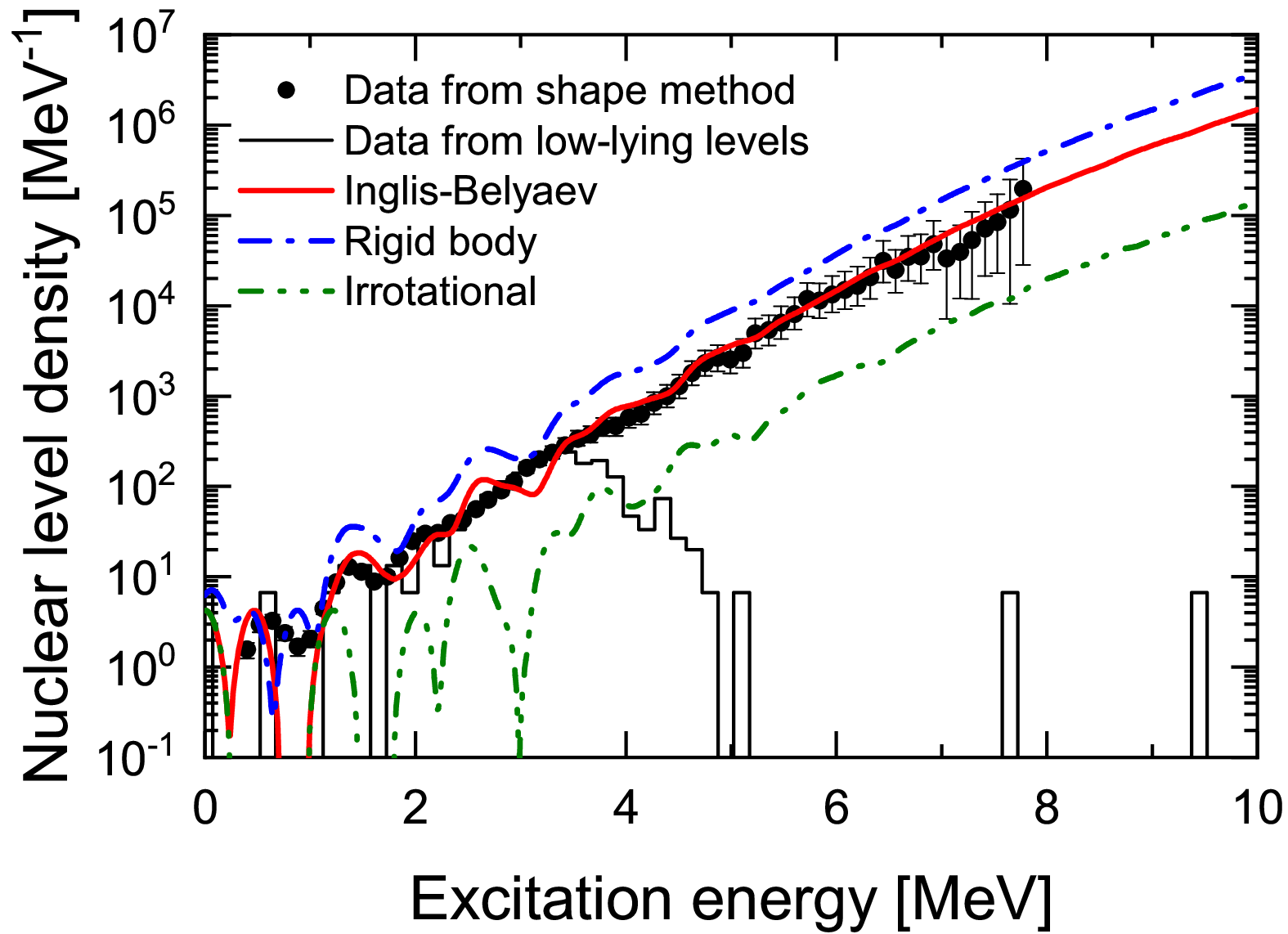}
  \caption{(color online) The nuclear level densities of $\mathrm{^{112}Cd}$ based on different formulas of moments of inertia.
  The data from shape method~\citep[]{Goriely2022PRC} are displayed with black dots and the data from low-lying levels in the RIPL3 database~\citep[]{RIPL3} are displayed with solid black line.}
  \label{fig3}
\end{figure}

The performances of different formulas of moments of inertia, i.e., rigid rotor, irrotational fluid, and the Inglis-Belyaev formula, in the nuclear level density prediction are compared in Fig.~\ref{fig3}, together with the data from shape method~\cite{Goriely2022PRC} and low-lying levels in RIPL3 database~\cite{RIPL3}.
The moments of inertia rotating around an axis perpendicular to the symmetry axis $\mathcal{J}_\perp$ calculated by these three formulas fulfill the relation $\mathcal{J}_\perp^\text{irr}<\mathcal{J}_\perp^\text{ing}<\mathcal{J}_\perp^\text{rig}$.
Naturally, a larger MOI would lead to larger nuclear level densities.
This is because a larger MOI leads to smaller rotational excitation energies as in Eq.~\eqref{rotenergy}, and thus leads to denser levels in rotation bands.
In this sense, the rigid rotor assumption may lead to too large of MOI and the irrotational fluid assumption may lead to too small of MOI in the calculations of nuclear level densities.
The Inglis-Belyaev formula provides a proper MOI and it reproduces the experimental data quite well.
This conclusion remains true even after taking into account the Thouless-Valatin dynamical rearrangement contributions by enhancing the Inglis-Belyaev MOI by 30\%~\cite{Libert1999PRC}.
Therefore, the Inglis-Belyaev formula is suggested to be used in the microscopic calculations of nuclear level densities and would be adopted in the following discussions.

\begin{figure}[!htbp]
  \centering
  \includegraphics[width=0.4\textwidth]{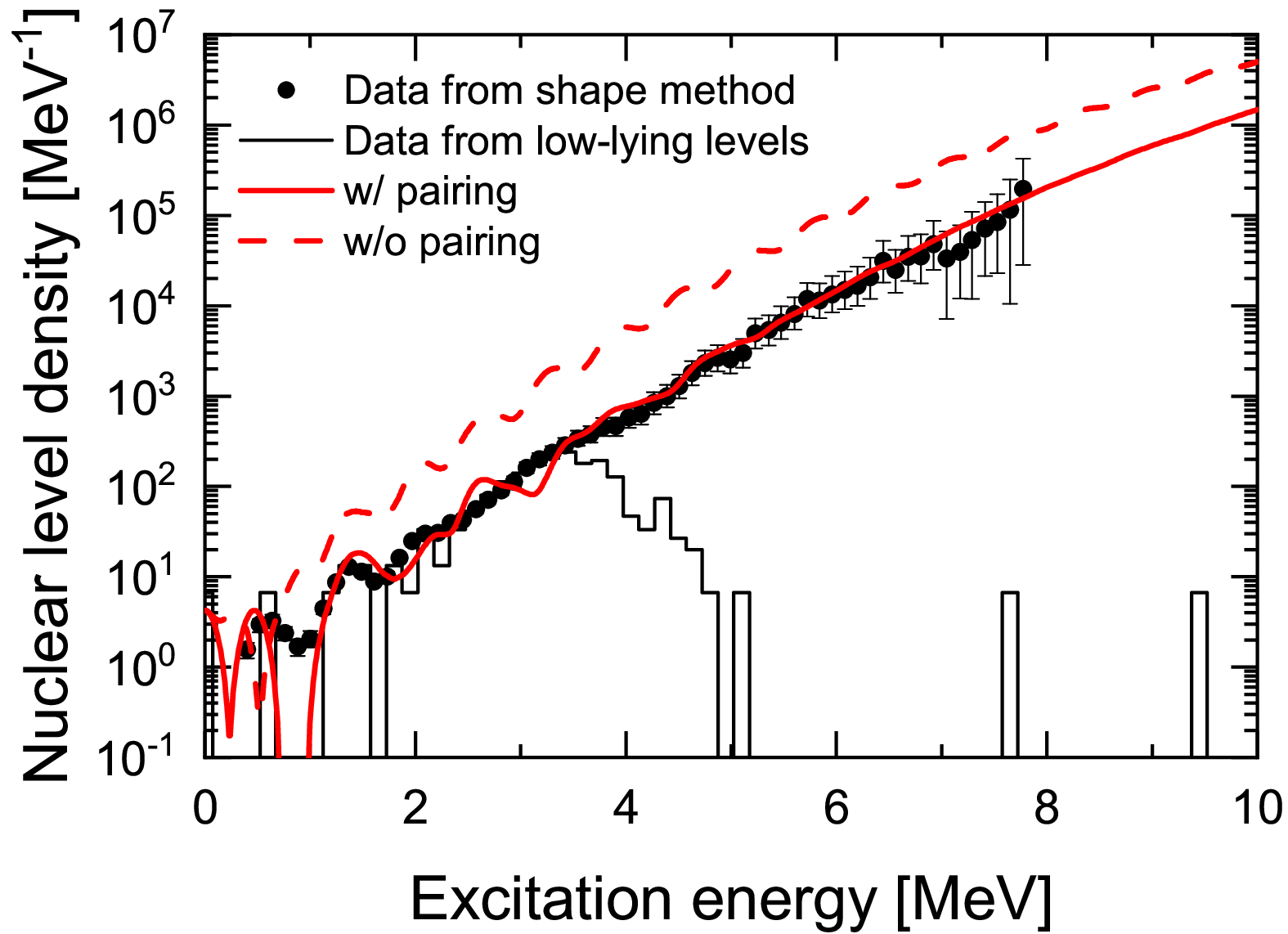}
  \caption{(color online) Comparison of the nuclear level densities of $\mathrm{^{112}Cd}$ calculated with (solid red line) and without pairing correlations (dashed red line).}
  \label{fig4}
\end{figure}

Pairing correlations are certainly important in the predictions of nuclear properties, which would certainly influence the predictions of nuclear level densities.
The nuclear level densities of $\mathrm{^{112}Cd}$ calculated with and without pairing correlations are presented in Fig.~\ref{fig4}.
It is found that the calculation with pairing correlations reproduces the experimental data for nuclear level densities quite well.
In contrast, the calculation without pairing correlations provides nuclear level densities of about one order of magnitude higher than the ones with pairing correlations.
The pairing correlations reduce the MOI and the density of incoherent ph states and, thus, lower the nuclear level density.

\begin{figure}[!htbp]
  \centering
  \includegraphics[width=0.4\textwidth]{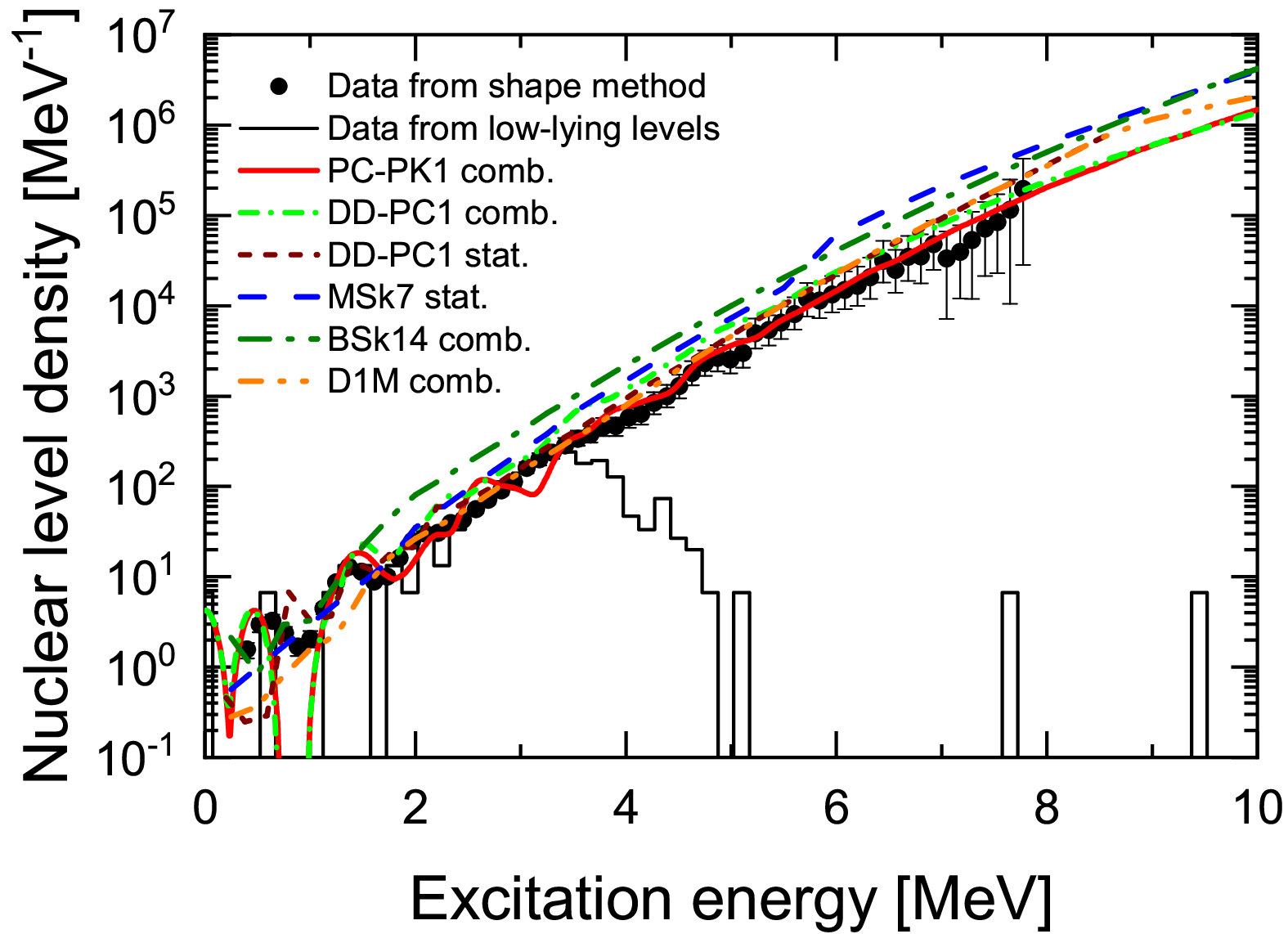}
  \caption{(Color online). The nuclear level densities of $\mathrm{^{112}Cd}$ calculated with combinatorial method based on RHB with PC-PK1 and DD-PC1, in comparison with the ones calculated with the statistical method based on Skyrme Hartree-Fock-BCS with MSk7~\cite{Demetriou2001NPA}, statistical method based on finite-temperature RHB with DD-PC1~\cite{Zhao2020PRC}, combinatorial method based on Skyrme Hartree-Fock-Bogoliubov with BSk14~\cite{Goriely2008PRC}, and combinatorial method based on temperature-dependent Gogny Hartree-Fock-Bogoliubov with D1M~\cite{Hilaire2012PRC}. The non-relativistic results are taken from TALYS code package~\cite{Koning2012NDS}.}
  \label{fig5}
\end{figure}

Finally, in Fig.~\ref{fig5}, the nuclear level densities of $\mathrm{^{112}Cd}$ calculated by the combinatorial method improved with Strutinsky method and based on RHB with PC-PK1 and DD-PC1 are compared with the ones calculated by the statistical method based on Skyrme Hartree-Fock-BCS with MSk7~\cite{Demetriou2001NPA}, statistical method based on finite-temperature RHB with DD-PC1~\cite{Zhao2020PRC}, combinatorial method based on Skyrme Hartree-Fock-Bogoliubov with BSk14~\cite{Goriely2008PRC}, and combinatorial method based on temperature-dependent Gogny Hartree-Fock-Bogoliubov with D1M~\cite{Hilaire2012PRC}.
The result based on PC-PK1 in this work not only reproduces the nuclear level density at high excitation energy, but also well describes the fluctuations at low excitation energy ($\lesssim2\ \mathrm{MeV}$).
With DD-PC1, both statistical method and combinatorial method predict slightly higher results at high excitation energy and similar fluctuations at low excitation energy. The excitation energy for the first peak predicted by statistical method with DD-PC1 is different from the ones by combinatorial method with DD-PC1 and PC-PK1 because the collective effects are taken into account in different ways.
In contrast, the D1M result well reproduces the experimental data at high excitation energy but fails at low excitation energy for the fluctuations.
The MSk7 and BSk14 results both significantly overestimate the nuclear level density at high excitation energy and also fail to describe the fluctuations.
The results in this work can be regarded as a benchmark of nuclear level density based on relativistic DFT, which well reproduces experimental data for nuclear level density of $\mathrm{^{112}Cd}$.

\section{Summary}\label{summary}

In summary, the nuclear level density is investigated with the combinatorial method based on relativistic density functional theory including pairing correlations for the first time.
In the combinatorial method, the Strutinsky method effectively removes the large fluctuations at low excitation energy and smoothes the total state density.
The microscopic Inglis-Belyaev formula provides a proper MOI and it reproduces the experimental data quite well.
The calculation with pairing correlations well reproduces the experimental data, while the calculation without pairing correlations provides nuclear level densities of about one order of magnitude higher.

Taking $\mathrm{^{112}Cd}$ as an example, it is demonstrated that the nuclear level density based on the relativistic density functional PC-PK1 can reproduce the experimental data at the same level as or even better than the previous approaches.
The successful descriptions of the nuclear level density with relativistic DFT pave a road for future developments of systematic calculations applied to the nuclear chart and further applications to neutron capture rates and \emph{r}-process.

\begin{acknowledgments}
  X.F.J thanks Yakun Wang and Yilong Yang for much help and stimulating discussions.
This work was partly supported by the National Natural Science Foundation of China (Grants No. 11935003, No. 11975031, No. 12141501, and No. 12070131001), the National Key R\&D Program of China (Contracts No. 2017YFE0116700 and No. 2018YFA0404400), the State Key Laboratory of Nuclear Physics and Technology, Peking University (Grants No. NPT2023ZX01 and No. NPT2023KFY02), and the High-performance Computing Platform of Peking University.
\end{acknowledgments}


%

\end{document}